     \documentstyle[proceedings,numreferences]{crckapb} 

\input epsf

\begin{opening}

\title{SINGLE-BUBBLE OPEN INFLATION: AN OVERVIEW}
\author{JUAN GARCIA-BELLIDO}
\institute{Theory Division, C.E.R.N.\\
           CH-1211 Gen\`eve 23, Switzerland}
\end{opening}

\runningtitle{OPEN INFLATION}

\begin{document}

\begin{abstract}
  The idea that the universe might be open is an old one, and the
  possibility of having an open universe arise form inflation is not
  new either. However, a concrete realization of a consistent
  single-bubble open inflation model is known only recently. There has
  been great progress in the last two years in the development of
  models of inflation consistent with observations in such an open
  universe.  In this overview I will describe the basic features and
  the phenomenological consequences of such models, making emphasis in
  the predictions of the CMB temperature anisotropies that differ from
  ordinary inflation.
\end{abstract}

\section{Introduction}

The idea that the universe might be open is an old one, see e.g.
\cite{Peebles}. Early attempts to accomodate standard inflation in an
open universe~\cite{ratra} failed to realize that in usual inflation
homogeneity implies flatness~\cite{turner}, due to the
Grishchuck-Zel'dovich effect~\cite{GZE}. The possibility of having a
truly open universe arise form inflation is not new either, see
\cite{Gott}, via the nucleation of a single bubble in de Sitter space.
However, a concrete realization of a consistent model is known only
recently, the single-bubble open inflation
model~\cite{singlebubble,LM}. Soon afterwards there was great progress
in determining the precise primordial spectra of perturbations~[8-19],
most of it based on quantum field theory in spatially open spaces.
Simultaneously there has been a large effort in model
building~\cite{LM,Green,induced,GBL} and constraining the existing models
from observations of the temperature power spectrum of cosmic
microwave background (CMB) anisotropies~[21-24].

In this review talk I will concentrate in model building and
constraints from CMB anisotropies. We will describe the nature of the
various primordial perturbations and give the corresponding spectra,
without deriving them from quantum field theoretical arguments. The
interested reader should find this in the literature. We will then
compute the corresponding angular power spectra of temperature
anisotropies in the CMB. Furthermore, we will give a review of the
different single- and mutiple-field open inflation models and
constrain their parameters from present observations of the CMB
anisotropies.  Sometimes this is enough to rule out some of the
models.  Finally, we will describe how future observations of the CMB
temperature and polarization anisotropies might be able to decide
among different inflationary models, both flat and open inflation
ones.

\section{Single-bubble open inflation}

The simplest inflationary model consistent with an open universe
arises from the nucleation of a bubble in de Sitter space~\cite{Gott},
inside which a second stage of inflation drives the spatial curvature
to {\it almost} flatness. The small deviation from flatness at the end
of inflation will be amplified by the subsequent expansion during the
radiation and matter eras. The present value of the density parameter
is determined from $|1-\Omega_0| \sim \exp(-2N_e)\times10^{56}$, where
$N_e\leq 65$ is the required number of e-folds during inflation
(inside the bubble), in order to give an open universe today. A few
percent change in $N_e$ could lead to an almost flat universe or a
wide open one.

This simple picture could be realized in the context of a single-field
scalar potential~\cite{singlebubble} or in multiple-field
potentials~\cite{LM,Green,induced,GBL}, as long as there exists a
false vacuum during which the universe becomes homogeneous and then
one of the fields tunnels to the true vacuum, creating a single
isolated bubble. The space-time inside this bubble is that of an open
universe \cite{CDL,Gott}. Although single-field models can in
principle be constructed, they require a certain amount of fine tuning
in order to avoid tunneling via the Hawking-Moss instanton~\cite{LM}.
The problem is that one needs a large mass for successful tunneling
and a small mass for successful slow-roll. For that reason, it seems
more natural to consider multiple-field models of open
inflation~\cite{LM,Green,induced,GBL}, where one field does the
tunneling, another drives slow-roll inflation inside the bubble, and
yet another may end inflation, like in the open hybrid model~\cite{GBL}.
Such models account for the large scale homogeneity observed by
COBE~\cite{COBE} and are also consistent with recent determinations of
a low density parameter~\cite{Dekel,Wendy,CB}.

The quantum tunneling can be described with the use of the bounce
action formalism developed by Coleman-DeLuccia~\cite{CDL} and
Parke~\cite{Parke} in the thin wall approximation, valid when the
width of the bubble wall is much smaller than the radius of curvature
of the bubble. This only requires that the barrier between the false
and the true vacuum be sufficiently high, $U_0 \gg \Delta U=U_F-U_T$.
In this case we can write the radius of the bubble in terms of
dimensionless parameters $a$ and $b$~\cite{JGB},
\begin{eqnarray}\label{RHT}
R_0 H_T &=& [1+(a+b)^2]^{-1/2} \equiv [1+\Delta^2]^{-1/2}\,,\\
&& \hspace*{-5mm} a\equiv{\Delta U\over3S_1H_T},\hspace{1cm} 
b\equiv{\kappa^2S_1\over4H_T}\,,\label{ab}
\end{eqnarray}
where $\kappa^2\equiv8\pi G$ and $S_1$ is the surface tension of the
bubble wall, which can be computed as $S_1 =
\int_{\sigma_F}^{\sigma_T} d\sigma\,[2(U(\sigma)- U_F)]^{1/2}$. Here
$\sigma$ is the tunneling field. Since $S_1\sim U_0/M \sim
M(\Delta\sigma)^2$ for a mass $M$ in the false vacuum, the parameter
$a\simeq (\Delta U/U_0)\,M/H_T$, which characterizes the degeneracy of
the vacua, can be made arbitrarily small by tuning $U_T\simeq U_F$. On
the other hand, the parameter $b\simeq (\Delta\sigma/M_{\rm Pl})^2
M/H_T$, which characterizes the width of the barrier, is not a tunable
parameter and could be very large or very small depending on the
model.

In order to prevent collisions with other nucleated bubbles (at least
in our past light cone) it is necessary that the probability of
tunneling be sufficiently suppressed. For an open universe of
$\Omega_0 > 0.2$, this is satisfied as long as the bounce action $S_B
> 6$, see Ref.~\cite{Gott}. This imposes only a very mild constraint
on the tunneling parameters $a$ and $b$, as long as the energy density
in the true vacuum satisfies $U_T \ll M_{\rm Pl}^4$, see
Ref.~\cite{JGB}.

\section{Primordial perturbation spectra}

There are essentially two kinds of primordial perturbations in open
inflation, quantum mechanical and classical or semiclassical. In the
first category we have the usual scalar and tensor metric
perturbations~\cite{sasaki,YST,TS}, modified in the context of an open
universe due to bubble nucleation, as well as
supercurvature~\cite{LW,super,YST} and bubble wall
perturbations~\cite{bubble,Garriga,Cohn,new}, which are specific to
open inflation. The other category, which appears in the context of
two-field open inflation~\cite{LM}, contains classical and
semiclassical effects due to tunneling to different values of the
inflaton field~\cite{slava,quasi}. All these perturbations create
anisotropies in the CMB which distort the angular power spectrum on
large scales (low multipoles). On smaller scales, at about one degree
separation (multipoles \ $l\sim200$), the (geometrical) effect of an
open universe is to shift the acoustic peaks of the temperature power
spectrum to higher multipoles~\cite{Mark}, but the primordial spectrum
there is essentially that of a flat universe.

\begin{figure}[t]
\centering
\hspace*{-6mm}
\leavevmode\epsfysize=4.2cm \epsfbox{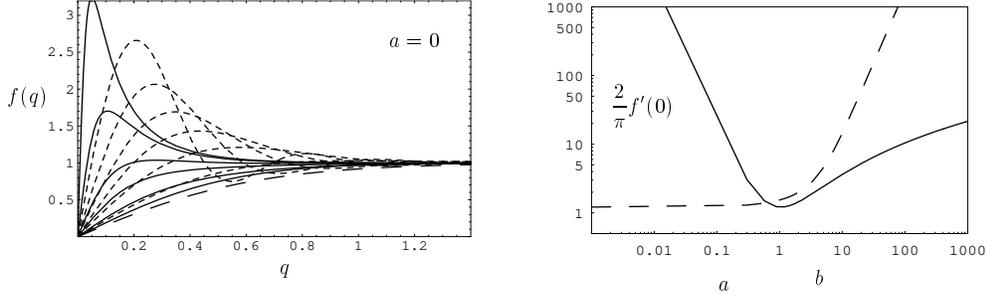}\\
\caption{\label{fig1} The left panel shows the spectral funtion $f(q)$
  in (\ref{fq}) as a function of $q$ for $a=0$ and
  $b=0.05,0.1,0.2,0.3,0.5,1.0$ (continuous lines, from top to bottom),
  and $b=2,5,10,20,50,200$ (dotted lines, from bottom to top). It is
  clear that $f(q)$ is linear in $q$ at $q=0$. The dashed line
  corresponds to $f(q)=\tanh\,\pi q/2$. \ The right panel shows the
  sharp increase in the slope $f'(q=0)$ as we increase or decrease $b$
  away from one (continuous line) and as we increase $a$ (dashed
  line). The minimum slope corresponds to $b\simeq1, a=0$, {\it i.e.}
  for almost degenerate vacua.}
\end{figure}

\vspace*{2mm}\noindent
{\it 3.1. Scalar and tensor perturbations}.\\[2mm]
Soon after the proposal of a single-bubble open inflation 
model~\cite{singlebubble}, the primordial spectrum of scalar 
perturbations was computed and found to be identical to the flat
case except for a prefactor that depends on the bubble 
geometry~\cite{YST},
\begin{equation}\label{AS2}
{\cal P}_{\cal R}(q) = A_S^2 \,f(q)\,, \hspace{1cm} 
A_S^2 = {\kappa^2\over2\epsilon} \Big({H_T\over2\pi}\Big)^2\,.
\end{equation}
The function $f(q)$ depends on the tunneling parameters $a$ and $b$,
see Eq.~(\ref{ab}),
\begin{equation}\label{fq}
f(q) = \coth\pi q - {z^2\cos\tilde q + 2q z \sin\tilde q\over
(4q^2 + z^2)\,\sinh\pi q }\,,
\end{equation}
where $\tilde q = q \ln((1+x)/(1-x))$ and
\begin{eqnarray}
x &=& (1-R_0^2H_T^2)^{1/2} = \Delta\,(1+\Delta^2)^{-1/2}\,, 
\label{x}\\[1mm]
z &=& (1-R_0^2H_T^2)^{1/2}-(1-R_0^2H_F^2)^{1/2} =
2b\,(1+\Delta^2)^{-1/2}\,, \label{z}
\end{eqnarray}
see Eq.~(\ref{RHT}). The function $f(q)$ is linear at small $q$, and
approaches a constant value $f(q) = 1$ at $q\geq2$, see
Fig.~\ref{fig1}. Here $q$ is the effective momentum in an open
universe, determined from $q^2=k^2-1$.  For scalar perturbations, the
effect of $f(q)$ on the temperature power spectrum is almost
negligible, and therefore the tilt of the scalar spectrum is
approximately given by the same formula as in flat space~\cite{LL93},
\begin{equation}\label{nS}
n_S - 1 \equiv {d\ln{\cal P}_{\cal R}(k)\over d\ln k} \simeq
- 6\epsilon + 2\eta \,,
\end{equation}
in the slow-roll approximation,
\begin{equation}\label{slowroll}
\epsilon = {1\over2\kappa^2}\,\Big({V'(\phi)\over V(\phi)}\Big)^2
\ll 1 \,, \hspace{1cm}
\eta = {1\over\kappa^2}\,\Big({V''(\phi)\over V(\phi)}\Big)
\ll 1 \,.
\end{equation}

On the other hand, the primordial spectrum of tensor or gravitational
waves' anisotropies took much longer to evaluate. There was for some
years the open problem that the observed power spectrum presented an
infrared divergence at $q=0$ in an open universe~\cite{Allen}. It was
clear that a physical regulator was necessary. But this is precisely
the role played by the bubble wall; recent computations have shown
that in the presence of the bubble the tensor primordial spectrum is
given by~\cite{TS}
\begin{equation}
{\cal P}_g(q) = A_T^2 \,f(q) \,, \hspace{1cm} \label{AT2}
A_T^2 = 8\kappa^2 \Big({H_T\over2\pi}\Big)^2\,.
\end{equation}
It is the shape of the function $f(q)$, see (\ref{fq}) and
Fig.~\ref{fig1}, which gives a finite physical observable, as I will
explain in the following section. Here $q$ is the effective momentum
for tensor modes, defined by $q^2=k^2-3$. While for scalar modes the
presence of this function $f(q)$ in the primordial spectrum becomes
irrelevant for observations, for tensor modes the slope of this
function at $q=0$ is an important ingredient in the final value of the
predicted power spectrum at low multipoles~\cite{JGB}.

\vspace*{2mm}\noindent
{\it 3.2. Supercurvature and bubble wall modes}.\\[2mm]
Apart from the usual continuum ($q^2\geq0$) of scalar and tensor
modes, generalized to an open universe, in single-bubble inflationary
models there is a new type of quantum fluctuations which are purely
geometrical, due to the boundary conditions associated with the
presence of the bubble wall in open de Sitter. These modes are
discrete modes, that appear whenever the mass of the scalar field in
the false vacuum is smaller than the rate of expansion, $m^2<2H^2$.

The supercurvature mode was first postulated in Ref.~\cite{LW} from
purely mathematical arguments related to homogeneous random fields.
Only later did concrete models of single-bubble open
inflation~\cite{sasaki} prove its existence in the primordial
spectrum of scalar fluctuations. Such a mode corresponds to an
infinite wavelength ($k^2=0, q^2=-1$) mode, and thus received its name
of {\it supercurvature} mode [the curvature scale corresponds to the
eigenmode with eigenvalue $k^2=1, q^2=0$]. These authors computed its
amplitude as~\cite{YST,super}
\begin{equation}\label{ASC}
A^2_{SC} = {\kappa^2\over2\epsilon}\,\Big({H_F\over2\pi}\Big)^2 =
A^2_S \, {H_F^2\over H_T^2} \,,
\end{equation}
where $A^2_S$ is given by Eq.~(\ref{AS2}).

However, this is not the only discrete supercurvature mode possible in
single-bubble models. As realized in Ref.~\cite{bubble}, there are
also scalar fluctuations of the bubble wall with $k^2=-3, q^2=-4$,
which could in principle have its imprint in the CMB insotropies. Such
modes were also discussed in Ref.~\cite{Hamazaki} and their amplitude
computed in Ref.~\cite{Garriga,Cohn}, based on field theoretical
arguments,
\begin{equation}\label{A2W}
A^2_W = {3\kappa^2\over2z}\,\Big({H_T\over2\pi}\Big)^2
= A_S^2\,{3\epsilon\over z}\,,
\end{equation}
where $z$ is given by Eq.~(\ref{z}) and $A^2_S$ is the scalar
amplitude (\ref{AS2}). Such modes contribute as transverse traceless
curvature perturbations~\cite{Garriga,bubble}, which nevertheless
behaves as a homogeneous random field~\cite{modes}. It was later 
realized~\cite{new} that the bubble wall fluctuation mode is actually
part of the tensor primordial spectrum, once gravitational backreaction
is included.

\vspace*{2mm}\noindent
{\it 3.3. Semiclassical perturbations and quasi-open inflation}.\\[2mm]
All single-field models of open inflation predict the above primordial
spectra of anisotropies: a continuum of scalar and tensor modes and
the discrete supercurvature and bubble wall modes. However, as
mentioned in the introduction, it is difficult to construct such
models without a certain amount of fine tuning~\cite{LM}, and thus
multiple-field models were considered~\cite{LM,Green,induced,GBL}.
However, a large class of two-field models do not lead to infinite
open universes, as it was previously thought, but to an ensemble of
very large but finite inflating `islands'. The reason is that the
quantum tunneling responsible for the nucleation of the bubble does
not occur simultaneously along both field directions and equal-time
hypersurfaces in the open universe are not synchronized with
equal-density or fixed-field hypersurfaces~\cite{LM,slava}. The most
probable tunneling trajectory corresponds to a value of the inflaton
field at the bottom of its potential; large values, necessary for the
second period of inflation inside the bubble, only arise as localized
fluctuations. The interior of each nucleated bubble will contain an
infinite number of such inflating regions of comoving size of order
$\gamma^{-1}$, where $\gamma$ depends on the parameters of the model,
see below. Each one of these islands will be a quasi-open universe. We
may happen to live in one of those patches of comoving size $d\leq
\gamma^{-1}$, where the universe appears to be open. This new effect,
semiclassical in origin, was recently discussed in Ref.~\cite{quasi}
and the amplitude of the associated fluctuation in the CMB was
computed as
\begin{equation}\label{AC}
A_C = {3\over2}{H_T^2\over m_T^2}\,\gamma \,, \hspace{1cm}
\gamma = {2\over3}{m_F^2\over H_F^2} + {1\over8}H_F^2R_0^4
(m_T^2-m_F^2)\,,
\end{equation}
where $R_0$ is the radius of the bubble at tunneling~(\ref{RHT}), and
the eigenvalue $\gamma$ was computed in the thin-wall
approximation~\cite{quasi}. As we shall see in the next section, many
of the present models are quasi-open. This does not mean that they are
not good cosmological models. If the co-moving size of the inflating
islands is sufficiently large, then the resulting classical anisotropy
may be unobservable.  This will prove rather constraining for the
models.

\begin{figure}[t]
\centering
\hspace*{-6mm}
\leavevmode\epsfysize=4cm \epsfbox{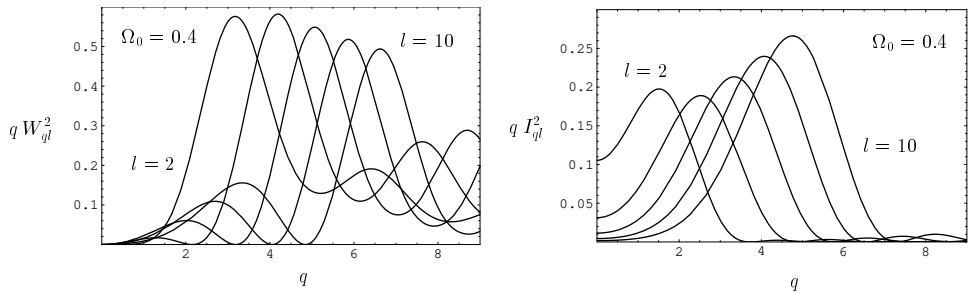}\\
\caption{\label{fig2} Windows functions for the scalar (left panel)
and tensor (right panel) CMB power spectra $l(l+1)C_l/A^2$, see
Eqs.~(\ref{CLS}) and (\ref{CLT}), for the first few multipoles,
$l=2,4,6,8,10$.}
\end{figure}

\section{CMB temperature anisotropies}

Quantum fluctuations of the inflaton field $\phi$ during inflation
produce long-wavelength scalar curvature perturbations and tensor
(gravitational waves) perturbations, which may leave their signature
in the CMB temperature aniso\-tropies, when they reenter the
horizon. Temperature anisotropies are usually given in terms of the
two-point correlation function or power spectrum, $C_l$, defined by an
expansion in multipole number $l$.  We are mainly interested in the
large scale (low multipole number) temperature anisotropies since it
is there where gravitational waves and the discrete modes could become
important. After $l\sim30$, the tensor power spectrum drops
down~\cite{Starobinsky,HW} while the density perturbation spectrum
increases towards the first acoustic peak, see Ref.~\cite{Silk}. On
these large scales the dominant effect is gravitational redshift via
the Sachs-Wolfe effect~\cite{SW67}. The scalar and tensor components
of the temperature power spectrum can be written as
\begin{eqnarray}\label{CLS}
l(l+1)C_l^S &=& \int_0^\infty dq\,A_S^2\,f(q)\,W_{ql}^2\,, \\[1mm]
l(l+1)C_l^T &=& \int_0^\infty dq\,A_T^2\,f(q)\,I_{ql}^2\,, \label{CLT}
\end{eqnarray}
where $W_{ql}$ and $I_{ql}$ are the corresponding window functions,
see Ref.~\cite{JGB}, which depend on the particular value of
$\Omega_0$. We have plotted these functions in Fig.~\ref{fig2} for a
typical value, $\Omega_0=0.4$, and the first few multipoles. Note that
the scalar window functions grow as a large power of $q$ at the
origin, so that the scalar power spectrum is rather insensitive to the
`hump' in the function $f(q)$, as mentioned above. On the other hand,
the tensor window functions are singular at $q=0$, and only the linear
dependence of $f(q)$ at the origin prevents the existence of the
infrared divergence found in \cite{Allen}. Furthermore, since the
functions $qI_{ql}^2$ are not negligible near the origin, the tensor
power spectrum turns out to be very sensitive to the `hump' in the
spectral function $f(q)$, see Fig.~\ref{fig1}.

\begin{figure}[t]
\centering
\hspace*{-4mm}
\leavevmode\epsfysize=8cm \epsfbox{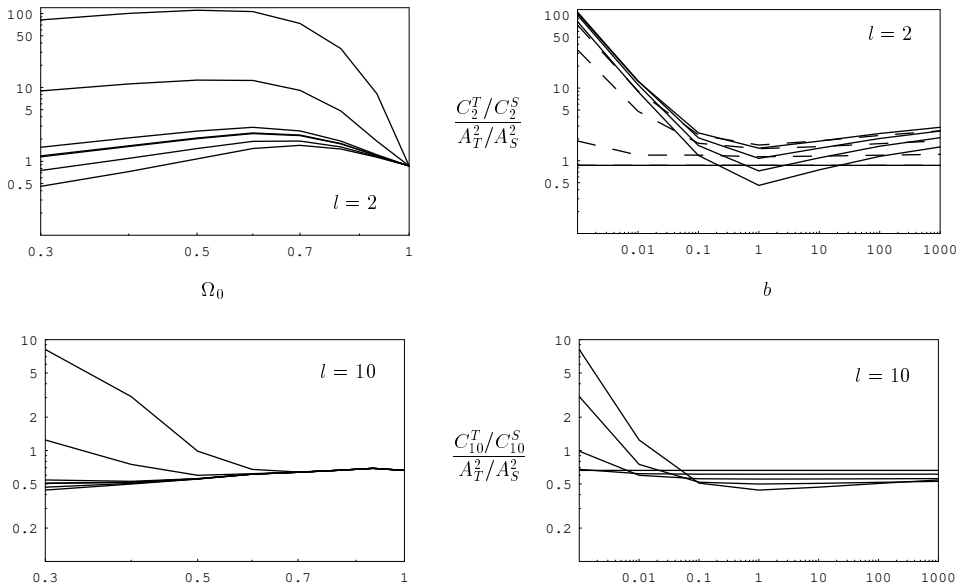}\\
\caption{\label{fig3} The ratio of tensor to scalar contribution to
  the CMB power spectrum $C_l^T/C_l^S$, normalized to the
  corresponding amplitudes, for the quadrupole and tenth multipole,
  for $a=0$, as a function of $\Omega_0$ (for $b=0.001, 0.01, 0.1, 1,
  10, 100, 1000$ from top to bottom) on the left panels, and as a
  function of the tunneling parameter $b$ (for $\Omega_0 = 0.3 - 1.0$) 
  on the right panels.}
\end{figure}

The ratio of tensor to scalar contribution to the temperature power
spectrum is a fundamental observable in standard inflation, which
depends on the slow-roll parameters (\ref{slowroll}) and provides a
consistency check of the theory~\cite{LL93}. In single-bubble open
inflation such a ratio depends not only on the slow-roll parameters
but also on the tunneling parameters (\ref{ab}) and on the value of
$\Omega_0$, see Ref.~\cite{JGB}, 
\begin{equation}\label{Oratio}
R_l = {C_l^T\over C_l^S} \simeq f_l(\Omega_0,a,b)\,
[1-1.3(n_S-1)]\,16\,\epsilon \,. 
\end{equation}
We have plotted this ratio in Fig.~\ref{fig3} for the quadrupole and
tenth multipole of the power spectrum, as a function of $\Omega_0$ and
the tunneling parameter $b$.

\begin{figure}[t]
\centering
\hspace*{-4mm}
\leavevmode\epsfysize=8.6cm \epsfbox{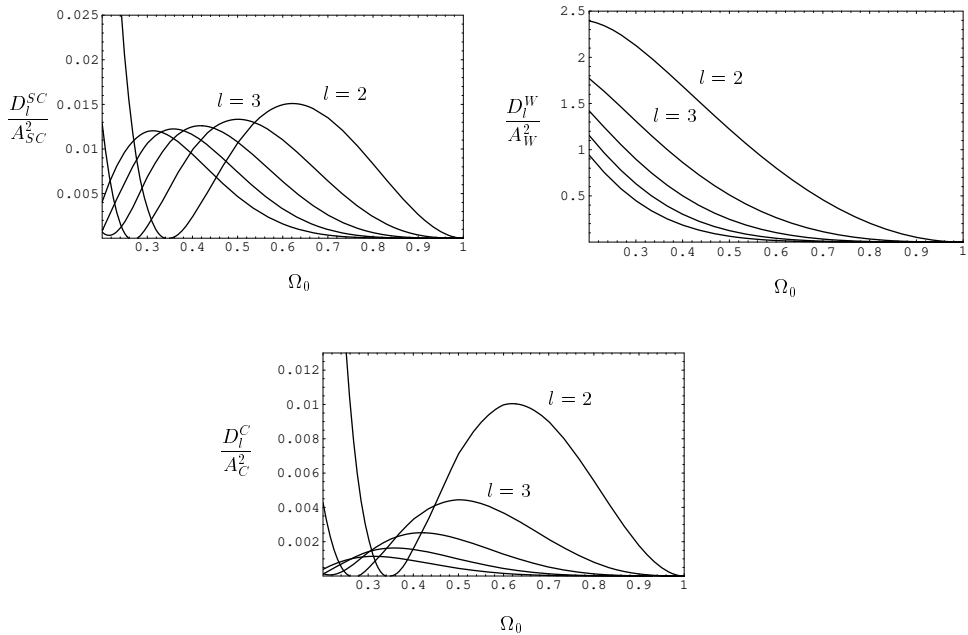}\\
\caption{\label{fig4} CMB power spectrum $l(l+1)\,C_l/A^2$, normalized
  to the corresponding amplitude, for the supercurvature, bubble wall
  and semiclassical fluctuations, as a function of $\Omega_0$, for the
  first few multipoles $l=2,3,...$ Note the dip in the supercurvature
  and semiclassical power spectra at different values of $\Omega_0$
  for different multipoles, due to accidental cancellations.}
\end{figure}

In the ideal case in which the gravitational wave perturbation can be
disentangled from the scalar component in future precise observations
of the CMB power spectrum~[38-42],
one might be
able to test this relation for a given value of $\Omega_0$.  This
would then constitute a check on the tunneling parameters $a$ and $b$.
Such prospects are however very bleak from measurements of the
temperature power spectrum alone, with the next generation of
satellites, see e.g.~\cite{Knox,COBRAS}. At most one can expect to
impose constraints on the parameters of the model from the absence of
a significant gravitational wave contribution to the CMB. However,
taking into account also the polarization power spectrum, together
with the temperature data, one expects to do much better, see
Refs.~\cite{SZ,ZSS} for the case of flat models. Hopefully similar
conclusions can be reached in the context of open models, and CMB
observations may be able to check the generalized consistency relation
(\ref{Oratio}) with some accuracy.

\begin{figure}[t]
\centering
\hspace*{-4mm}
\leavevmode\epsfysize=4.3cm \epsfbox{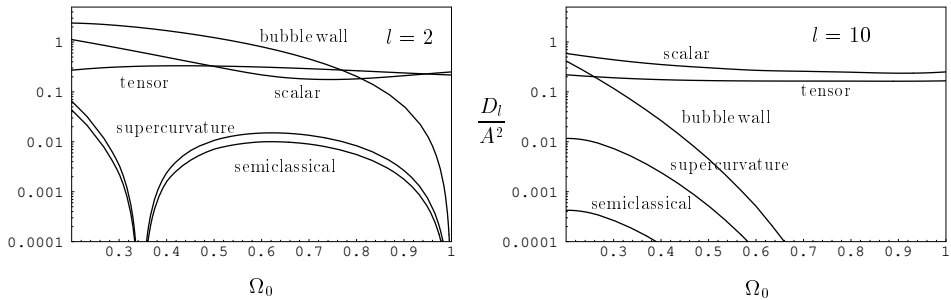}\\
\caption{\label{fig5} Quadrupole (left panel) and tenth multipole
  (right panel) of the CMB power spectra, normalized to the
  corresponding amplitude, $l(l+1)\,C_l/A^2$, for the tensor, scalar,
  supercurvature, bubble wall and semiclassical primordial spectra.
  We are assuming the minimal contribution from tensors ($a=0, b=1$).}
\end{figure}

The supercurvature, bubble wall and semiclassical modes also
contribute to the temperature power spectrum. Their contribution is
plotted in Fig.~\ref{fig4}, normalized to the corresponding amplitude,
see Eqs.~(\ref{ASC}), (\ref{A2W}) and (\ref{AC}), as a function of
$\Omega_0$ for the first few multipoles. Note the dip in the spectrum
at certain values of $\Omega_0$ due to accidental
cancellations~\cite{induced,GBL,JGB}. This does not affect the bounds
since higher multipoles will fill in those gaps. The relative
importance of the different components of the power spectrum is
crucial in order to derive bounds on the model parameters.  We have
plotted in Fig.~\ref{fig5} the quadrupole and tenth multipole of the
CMB power spectrum for each mode, normalized to their corresponding
amplitudes. Assuming that the observed quadrupole only comes from the
scalar component, one can deduce the following constraints, see
Ref.~\cite{JGB,quasi},
\begin{eqnarray}\label{bounds}
H_T&=&\sqrt{\pi\epsilon}\ 5\times 10^{-5} M_{\rm Pl}\,,\\ 
\label{super} H_F &<& 10 \,H_T \,,\\ 
\label{bubble} \epsilon &<& z/3 \,,\\
\label{gamma} \gamma &<& {m_T^2\over H_T^2}\ 3\times 10^{-4}\,.
\end{eqnarray}
The third expression accounts for both the tensor and the bubble wall
constraints, since the bubble wall fluctuation is actually part of the
tensor spectrum~\cite{new} and gives the largest contribution at low
multipoles. The constraints coming from higher multipoles are
significantly weaker, see Fig.~\ref{fig5}. One could argue that the
quadrupole is going to be hidden in the cosmic variance~\cite{BET} and
thus only the constraints at higher multipoles, say $l=10$, should be
imposed.  However, polarization power spectra may one day be used to
get around cosmic variance~\cite{Loeb} and we may be able to extract
information about the scalar and tensor components at low multipoles.
We will therefore take the conservative attitude that consistent
models of open inflation should satisfy the bounds coming from the
quadrupole, (\ref{bounds})-(\ref{gamma}).

\section{Model building}

In this section we will review the different single-bubble open
inflation models present in the literature, and use the CMB
observations to rule out some of them and severely constrain others.
We will see that open inflation models could be as predictive as
ordinary inflation, in the sense that they also can be ruled out
if they are in conflict with observations.

As mentioned in the introduction, single-field models of open
inflation~\cite{singlebubble} require some finetuning in order to have
a large mass for successfull tunneling and a small mass for slow-roll
inside the bubble~\cite{LM}.  Even if such a model can be constructed
from particle physics, it still needs to satisfy the constraints
coming from observations of the CMB anisotropies. These models lack
both supercurvature and semiclassical anisotropies, by construction.
However, they generically produce too large tensor modes at low
multipoles, where it is dominated by the bubble wall fluctuations.
The reason is that in these models the tunneling parameter $b$ that
characterizes the width of the tunneling barrier~(\ref{ab}) is
extremely small and it is very difficult to satisfy the
bound~(\ref{bubble}) due to the bubble wall fluctuations, unless one
does extreme finetuning of the parameters.  Let us analyze a typical
example, which is a variant of the new inflation
type~\cite{singlebubble}. Tunneling occurs from a symmetric phase at
$\sigma=0$ to a value $\sigma_b$ from which the field slowly rolls
down the potential towards the symmetry breaking phase at
$\sigma=v\sim M_{\rm GUT}\sim 10^{15}$ GeV. The fact that we have a
finite number of e-folds, $N=60$, requires $\sigma_b\sim
v\,\exp(-\alpha N)\ll v$, where $\alpha\simeq 2m_T^2/3H_T^2$. The rate
of expansion in the false vacuum is of order that in the true vacuum,
$H_T=(8\pi V(0)/3M_{\rm Pl}^2)^{1/2} \sim 3\times10^{-6} M_{\rm Pl}$,
which implies $\epsilon\sim10^{-3}$. Taking a typical mass in the
false vacuum to be $M\sim M_{\rm GUT}$, we find
\begin{equation}
b_{\rm single} \simeq \Big({\sigma_b\over M_{\rm Pl}}\Big)^2 {M\over H_T} \sim
5\times10^{-9}\,,
\end{equation}
which gives $z\sim 2b \sim 10^{-8}$, an extremely small number
that makes it impossible to satisfy the bubble wall constraint,
$\epsilon < z/3$, see Eq.~(\ref{bubble}). 

In other words, the simplest single-field models of open
inflation~\cite{singlebubble} are not only fine tuned but actually
produce too large gravitational wave anisotropies in the CMB on large
scales to be consistent with observations. Perhaps more complicated
models could still be fine tuned to agree with observations.

\vspace*{2mm}\noindent
{\it 5.1. Coupled and uncoupled two-field models}.\\[2mm]
In this section we shall consider a class of two-field
models~\cite{LM} with a potential of the form
\begin{equation}
V(\sigma, \phi)= V_0(\sigma)+{1\over 2}m^2\phi^2 + 
{1\over 2} g^2 \sigma^2\phi^2.
\label{general}
\end{equation}
Here $V_0$ is a non-degenerate double well potential, with a false
vacuum at $\sigma=0$ and a true vacuum at $\sigma=v$.  When $\sigma$
is in the false vacuum, $V_0$ dominates the energy density and we have
an initial de Sitter phase with expansion rate given by $H_F^2\approx
8\pi V_0(0)/3M_{\rm Pl}^2$. Once a bubble of true vacuum $\sigma=v$
forms, the energy density of the slow-roll field $\phi$ may drive a
second period of inflation. However, as pointed out in Ref.~\cite{LM},
the simplest two-field model of open inflation, given by
(\ref{general}) with $g=0$ and $m\neq 0$, is actually a quasi-open one
because equal-time hypersurfaces, defined by the $\sigma$ field after
nucleation, are not synchronized with equal-density hypersurfaces,
determined by the slow-roll of the $\phi$ field during inflation
inside the bubble. In order to suppress this effect it was
argued~\cite{LM} that a large rate of expansion in the false vacuum
compared to the true vacuum, $H_F\gg H_T$, could prevent the $\phi$
field from rolling outside the bubble and distorting the equal-density
hypersurfaces inside the bubble.  However, this would
induce~\cite{super} a large supercurvature mode anisotropy in the CMB,
which would be incompatible with observations. A careful
analysis~\cite{slava,quasi} shows that indeed the effect is important
at low multipoles. In this model, $m_F=m_T=m$, the supercurvature
eigenvalue $\gamma = 2m_F^2/3H_F^2$, and in order to satisfy
(\ref{gamma}) it requires $H_F > 50 H_T$, which is incompatible with
the supercurvature constraint, $H_F < 10 H_T$.  Therefore, this model
seems to be ruled out.

\begin{figure}[t]
\centering
\hspace*{-5mm}
\leavevmode\epsfysize=8.5cm \epsfbox{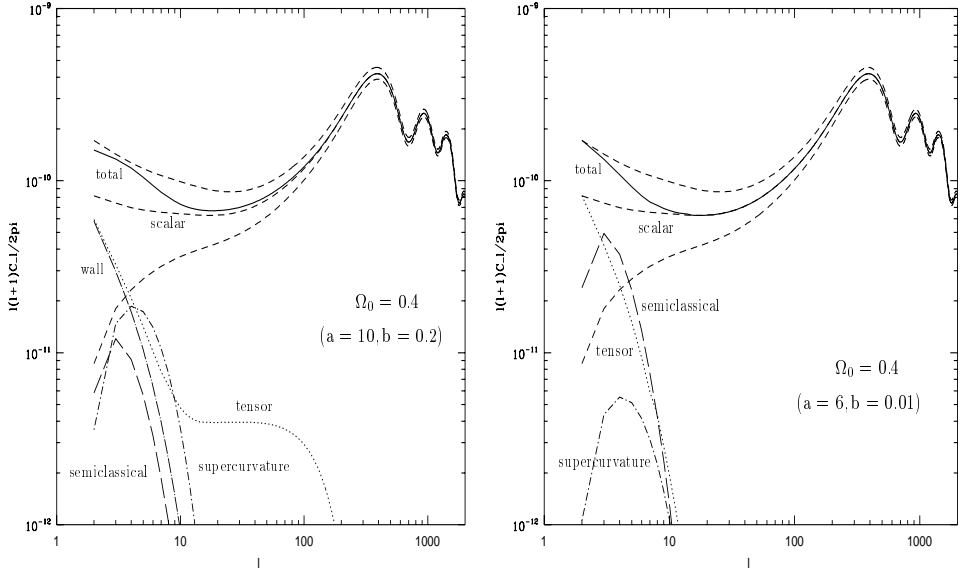}\\
\caption{\label{fig6} The complete angular power spectrum of
temperature anisotropies for the coupled two-field model (left panel)
for $\Omega_0 = 0.4$ and $(a=10, b=0.2)$, and the open hybrid model
(right panel) for $\Omega_0 = 0.4$ and $(a=6, b=0.01)$. We show also
the individual contributions from the scalar, tensor, supercurvature,
semiclassical and bubble wall modes, as well as the expected cosmic
variance (dashed lines) for 1/3 of the sky coverage from the future
CMB satellites. Note that the bubble wall mode is responsible for the
large growth of the tensor contribution at low multipoles. Only the
scalar modes remain beyond about $l=50$, where they grow towards the
acoustic peaks.}
\end{figure}

In order to construct a truly open model, Linde and Mezhlumian
suggested taking $m=0$ and $g\neq 0$, i.e. the ``coupled'' two-field
model~\cite{LM}. In this way, the mass of the slow-roll field vanishes
in the false vacuum, and it would appear that the problem of classical
evolution outside the bubble is circumvented.  However, as we showed
in Ref.~\cite{quasi}, this is not exactly so, and actually the whole
class of models (\ref{general}) leads to quasi-open universes, which
are constrained by CMB observations. Let us work out those constraints
in detail. We will assume a tunneling potential like~\cite{LM}
\begin{equation}\label{tunpot}
V_0(\sigma)= V_0+{1\over 2}M^2\sigma^2 - \alpha M\sigma^3 +
{1\over 4} \lambda \sigma^4\,.
\end{equation}
For $\alpha=\sqrt\lambda$ we have $\sigma_c=2M/\sqrt\lambda$ and
$\phi_c=M/g$. The field can tunnel for $\phi<\phi_c$. The constant
$V_0\simeq2.77 M^4/\lambda$ has been added to ensure that the absolute
minimum, at $\phi=0$ and $\sigma_0\simeq1.3\,\sigma_c$, has vanishing
cosmological constant. After tunneling, the field $\phi$ moves along
an effective potential $V(\phi)=m^2\phi^2/2$, where the effective mass
varies only slightly from tunneling to the end of inflation,
$m\simeq1.3\,g\sigma_c$. This potential drives a period of chaotic
inflation with slow-roll parameters $\epsilon=\eta=1/2N_e=1/120$.
Substituting into (\ref{bounds}) we find $H_T = 6.32\,m = 8\times
10^{-6} M_{\rm Pl}$, and therefore $g\sigma_c = 10^{-6} M_{\rm
  Pl}$. The rate of expansion in the false vacuum is determined from
$H_F^2/H_T^2 = 1 + 4ab$, where $b = (4\pi\sqrt2/3\lambda) M^3/H_T
M_{\rm Pl}^2$, which gives
\begin{equation}\label{MHF}
M = {(1+4ab)^{1/2}\over4b}\,H_F \geq \sqrt2 H_F\,,
\end{equation}
the last condition arising from preventing the formation of the bubble
through the Hawking-Moss instanton, see Ref.~\cite{LM}. Furthermore,
taking $m_F=0$ in the eigenvalue $\gamma$ gives (\ref{AC}) 
\begin{equation}\label{ACN}
A_C = {3\over16}{H_F^2\over H_T^2} (R_0H_T)^4 = 
{3(1+4ab)\over16[1+(a+b)^2]^2} < 4\times 10^{-4}\,.
\end{equation}
From the supercurvature mode condition (\ref{super}), $(1+4ab)^{1/2} <
10$, together with (\ref{MHF}) we find the constraint $b < 1$. From
Eq.~(\ref{ACN}), we realize that having nearly degenerate vacua,
$a\ll1$, is not compatible with observations.  Satisfying (\ref{ACN})
would require $b\ll1$ and $a\gg1$. However, for these values of the
parameters we expect large tensor contributions, see Fig.~\ref{fig3}.
So there should be a compromise between the different mode
contributions. For details, see Ref.~\cite{CMBquasi}.

We have shown in the left panel of Fig.~\ref{fig6}a the complete
temperature power spectrum for a coupled two-field model having $a=10$
and $b=0.2$, for $\Omega_0=0.4$, which is consistent with
observations. It has contributions from all the modes: scalar, tensor,
supercurvature, semiclassical and bubble wall.  Note however that the
bubble wall mode is in fact included in the sharp growth of the tensor
contribution at small multipole number, as emphasized in
Ref.~\cite{new} and shown explicitly in Fig.~\ref{fig6}a, and should
not be counted twice. Although it is in principle possible to
construct a model consistent with observations, the parameters of the
model are not very natural. In this simple model (\ref{tunpot}) one
would expect that quantum tunneling would occur soon after the
critical point, $\phi=\phi_c$, where $a\ll1$ and $b\sim1$. In that
case, instead of an infinite open universe we would actually live in a
finite quasi-open one~\cite{quasi}. In order to suppress the
associated semiclassical anisotropy we had to choose other values of
the parameters. As can be seen from Fig.~\ref{fig6}a, there exists a
range of parameters for which all contributions to the CMB
anisotropies are compatible with observations.

\vspace*{2mm}\noindent
{\it 5.2. Supernatural open inflation}.\\[2mm]
An attractive scenario for open inflation is the model of a complex
scalar field with a slightly tilted mexican hat potential, where the
radial component of the field does the tunneling and the
pseudo-Goldstone mode does the slow-roll. This model was called
``supernatural'' inflation in Ref.~\cite{LM}, because the hierarchy
between tunneling and slow-roll mass scales is protected by the
approximate global $U(1)$ symmetry. Expanding the field in the form
$\Phi = (\sigma/\sqrt2)\exp(i\phi/v)$, where $v$ is the expectation
value of $\sigma$ in the broken phase, we consider a potential of the
form $V=V_0(\sigma) + V_1(\sigma,\phi)$, where $V_0$ is $U(1)$
invariant and $V_1$ is a small perturbation that breaks this
invariance. It is assumed that $V$ has a local minimum at $\Phi=0$
which makes the symmetric phase metastable.  We shall consider a tilt
in the potential of the form $V_1= \Lambda^4(\sigma)G(\phi)$ where
$\Lambda$ is a slowly varying function of $\sigma$ which vanishes at
$\sigma=0$. For definiteness we can take $G=(1-\cos\phi/v)$. The idea
is that $\sigma$ tunnels from the symmetric phase $\sigma=0$ to the
broken phase $\sigma_0=v$, landing at a certain value of $\phi$ away
from the minimum of the tilted bottom. Once in the broken phase, the
potential $V_1$ cannot be neglected, and the field $\phi$ slowly rolls
down to its minimum, driving a second period of inflation inside the
bubble.  An attractive feature of this model is that depending on the
value of $\phi$ on which we end after tunneling, the number of
$e$-foldings of inflation will be different. Hence it appears that in
principle we can get a different value of the density parameter in
each nucleated bubble. As we shall see, however, the picture is
somewhat different. For $V_1(\phi)=\Lambda^4(1-\cos\phi/v)$ we find
the slow-roll parameters $\epsilon=(1/2\kappa^2v^2)\cot^2\phi/2v \ll
1$ and $\eta = \epsilon - 1/2\kappa^2v^2$. From the constraint on the
spectral tilt, $n_S-1=-4\epsilon-1/\kappa^2v^2 > -0.2$, we find that
necessarily $\kappa^2v^2 > 5$, which means that the vev of $\sigma$ is
$v\simeq M_{\rm Pl}$. We are again in a situation similar to the
single-field models, where we need some extreme fine tuning to prevent
the Hawking-Moss instanton from forming the bubble, see
Ref.~\cite{LM}. Indeed, for a generic tunneling potential like
(\ref{tunpot}) we have $V_0\simeq M^2 \sigma_0^2/2$ and thus
$H_F\simeq 2M\,\sigma_0/M_{\rm Pl}\geq M$.  Under this conditions the
tunneling does {\em not} occur along the Coleman-DeLuccia instanton,
which is necessary for the formation of an open universe inside the
bubble. The only way to prevent this is by artificially bending the
potential so that it has a large mass at the false vacuum. In
Ref.~\cite{LM} it was proposed a way to lower the minimum at the
center of the Mexican hat via radiative corrections from a coupling of
the $U(1)$ field $\Phi$ to another scalar $\chi$. For certain values 
of the coupling constant, $g^4=32\pi\lambda$, it is possible to make the
two minima, at $\sigma=0$ and $\sigma_0$, exactly degenerate. The
tunneling potential is then
\begin{equation}\label{tunpot2}
V_0(\sigma)= {\lambda\over 2}(\sigma_0^2-\sigma^2)\sigma^2 +
\lambda \sigma^4\,\ln{\sigma\over\sigma_0}\,,
\end{equation}
where $\sigma_0=M/\sqrt\lambda\simeq M_{\rm Pl}$. The associated
tunneling parameters become $a=0$ and $b=(\sigma_0/M_{\rm Pl})^2M/H_T
\simeq M/H_T$, which can be large. As emphasized in Ref.~\cite{quasi}
there is a supercurvature mode in this model, associated with the
massless Goldstone mode, which induces both supercurvature and
semiclassical perturbations. Because of the different normalization of
the supercurvature mode in supernatural inflation, $H_F \to 2R_0^{-1}$,
the supercurvature constraint~(\ref{super}) should read in this case
\begin{equation}\label{super2}
R_0H_T > 0.2\,,
\end{equation}
which is not trivially satisfied. On the other hand, the eigenvalue
$\gamma=R_0^2m_T^2/2$ for the Goldstone mode~\cite{quasi} induces a
large semiclassical perturbation (\ref{gamma}) unless 
\begin{equation}\label{class2}
R_0H_T < 0.024\,.
\end{equation}
It is clear that these two constraints cannot be accommodated
simultaneously, and thus the model is incompatible with observations.
One still has to make sure~\cite{CMBquasi} that a slight modification
of the potential does not give a certain range of parameters in which
the model works.

\vspace*{2mm}
\noindent 
{\it 5.3. Induced gravity open inflation}.\\[2mm] 
This model was proposed in Ref.~\cite{Green} as a way of avoiding the
problems of classical motion outside the bubble. In the false vacuum
the inflaton field is trapped due to its non-minimal coupling to
gravity, with coupling $\xi$. When the tunneling occurs it is left
free to slide down its symmetry breaking potential $V(\varphi) =
\lambda(\varphi^2-\nu^2)^2/8$. The expectation value of the inflaton
at the global minimum gives the Planck mass today, $M_{\rm Pl}^2 =
8\pi\xi\nu^2$. The model is parametrized by $\alpha =
8U_F/\lambda\nu^4$, which determines the value of the stable fixed
point in the false vacuum, $\varphi_{\rm st}^2 = \nu^2 (1+\alpha)$, 
as well as the difference in rates of expansion in the false and true
vacua, $H_F^2=H_T^2\,(1+\alpha)/\alpha$, and the slow-roll parameters,
$\epsilon = 8\xi/(1+6\xi)\alpha^2$, $\eta =
8\xi\,(1-\alpha)/(1+6\xi)\alpha^2$, see Ref.~\cite{induced}. 

We will assume a tunneling potential for the $\sigma$ field of the type
\begin{equation}
U(\sigma) = {1\over4}\lambda'\sigma^2(\sigma-\sigma_0)^2 + \mu U_0
\Big[1-\Big({\sigma\over\sigma_0}\Big)^4\Big]\,,
\end{equation}
where $\sigma_0 = M\sqrt{2/\lambda}$, $U_0=M^4/4\lambda'$ and
$\mu\ll1$ for the thin wall approximation to be valid. This form of
the potential gives a tunneling parameter $b = (2\pi/3\lambda) M^3/H_T
M_{\rm Pl}^2$, which determines the relation between the mass of the
$\sigma$ field in the false vacuum and the rate of expansion there, $M
= H_F\,(1+1/\alpha)^{1/2}/\mu b$. Thanks to $\mu\ll1$, we can have
$M\gg H_F$ for values of $b\geq1$, which induces gravitational wave
anisotropies that are well under control.

Furthermore, the induced gravity model seems to be truly open, since
the inflaton field $\varphi$ is static in the false vacuum and thus
there is no supercurvature mode associated with classical motion
outside the bubble, see Ref.~\cite{quasi}. Therefore the
constrain~(\ref{gamma}) does not apply, and there exists for this
model a range of parameters for which all contributions to the CMB
anisotropies are compatible with observations, see Ref.~\cite{JGB}.
However, the instanton may not take you to $\varphi_{\rm st}$ in the
true vacuum, but to a different value, closer to the minimum of the
potential, $\varphi=v$. In that case, the number of $e$-folds is
smaller than expected and so is the value of $\Omega_0$. Such effects
should be taken into account for the determination of the model
parameters, see Ref.~\cite{CMBquasi}.

\vspace*{2mm}
\noindent 
{\it 5.4. Open hybrid inflation}.\\[2mm] 
This model was proposed recently~\cite{GBL} in an attempt to produce a
significantly tilted scalar spectrum in the context of open inflation
to be in agreement with large scale structure~\cite{WS}. It is based
on the hybrid inflation scenario~\cite{hybrid,CLLSW}, which has
recently received some attention from the point of view of particle
physics~[50-54],
together with a tunneling
field which sets the initial conditions inside the bubble.

In this model there are three fields: the tunneling field $\sigma$,
the inflaton field $\phi$ and the triggering field $\psi$. The
tunneling occurs like in the coupled model of section 5.1 with
potential
\begin{equation}\label{pot}
U(\sigma,\phi)= V_0+{\lambda\over4}\sigma^2(\sigma-\sigma_0)^2
+ {1\over2}g^2(\phi^2-\phi_c^2)\sigma^2 + U_0\,,
\end{equation}
where $\sigma_0 = 2M/\sqrt\lambda$, \,$\phi_c = M/g$, \,$V_0\simeq
2.77M^4/\lambda$ to ensure that at the global minimum we have
vanishing cosmological constant, and $U_0$ is the vaccum energy
density associated with the triggering field. We satisfy $V_0\ll
U_0$. If the $\sigma$ field tunnels when $\phi=3\phi_c/4$, then
$\Delta U = U_F-U_T \simeq V_0/2 \simeq m^2\phi_c^2/4$. After that
the inflaton field will slow-roll down the effective potential
$U=U_0+m^2\phi^2/2\simeq U_0$ driving hybrid inflation, until the
coupling to $\psi$ triggers its end. The model is parametrized by
$\alpha = m^2/H^2$, see Refs.~\cite{GBL,JGB,CMBquasi}, in terms of
which the spectral tilts can be written as $n_S-1=2\alpha/3-2\epsilon$
and $n_T=-2\epsilon$. At tunneling we can write $V_0\simeq
m^2\phi_c^2/2=8m^2\phi_T^2/9$, so that the slow-roll parameter 
$\epsilon = (\alpha/3) 9V_0/16U_T = 3\alpha ab/2$, where $4ab = 
\Delta U/U_T\simeq V_0/2U_T$, see (\ref{ab}), and thus 
\begin{equation}\label{ns}
n_S = 1 + {2\alpha\over3}\,\Big(1 - {9ab\over2}\Big).
\end{equation}
In order for open hybrid to have a large tilt we require {\em both}
a large $\alpha$ and a small value of $ab$. As we will show, this is
not going to be possible due to the set of conditions 
(\ref{bounds}-\ref{gamma}) from the CMB. 

For that purpose, we should compute the tunneling parameter $b =
(4\pi\sqrt2/3\lambda) M^3/H_T M_{\rm Pl}^2 \simeq (V_0/4U_T)\,H_T/M$,
which requires $M=2a H_T>2H_T$, and thus $a>1$. Since both $H_T<M$ and
$V_0\ll U_T$, we expect $b\ll1$, which will induce large tensor
anisotropies at low multipoles. This is a generic feature of open
hybrid models. The tensor amplitude is related to the scalar one by
$A_T^2 = 16\epsilon A_S^2 = 8A_S^2\,(n_S-1)\,9ab/(2-9ab)$. In order to
satisfy the CMB constraints we require
\begin{eqnarray}
H_F^2/H_T^2 &=& 1 + 4ab < 100 \,,\\ 
\epsilon &=& {3\alpha a b\over2} < {2b\over3[1+(a+b)^2]^{1/2}} \,,\\
A_C &=& {3(1+4ab)\over16[1+(a+b)^2]^2} < 4\times 10^{-4}\,,\\
M &=& {2a\over(1+4ab)^{1/2}} H_F > 2 H_F\,.
\end{eqnarray}
Since $a>1$ requires $b<V_0/8U_T\ll1$, we can use the third constraint
to get the bound $a>5$, which then imposes (through the second
constraint) that $\alpha < 4/9a(1+a^2)^{1/2}\simeq 1/2a^2 <
1/50$. This means that the scalar tilt (\ref{ns}) cannot be
significantly larger than 1, as was the aim of Ref.~\cite{GBL}.

We have plotted in Fig.~\ref{fig6}b the complete angular power
spectrum of temperature anisotropies for the open hybrid model, in the
case $\Omega_0=0.4$ and $(a=6, b=0.01)$. We have also included the
cosmic variance uncertainty~\cite{JKKS}
\begin{equation}\label{cosmic}
\sigma_l = \Big[{2\over(2l+1)f_{\rm sky}}\Big]^{1/2}\, C_l\,,
\end{equation}
where we have chosen $f_{\rm sky}=1/3$ as the typical fraction of the
sky covered by the future satellites MAP~\cite{MAP} and
PLANCK~\cite{COBRAS}. In order to prevent the tensor contribution from
exceeding the cosmic variance we had to reduce the scalar spectral
tilt to $n=1.006$, which is essentially scale invariant and may not be
enough to allow consistency with large scale structure~\cite{WS}.
Furthermore, as we decrease in $\Omega_0$ it will be necessary to have
scalar spectra which are closer and closer to scale invariance, in
order to reduce the tensor contribution. In any case, there exists for
this model a range of parameters for which all contributions to the
CMB anisotropies, see e.g. Fig.~\ref{fig6}b, are compatible with
observations.

\section{Conclusions}

Single-bubble open inflation is an ingenious way of reconciling an
infinite open universe with the inflationary paradigm. In this
scenario, a symmetric bubble nucleates in de Sitter space and its
interior undergoes a second stage of slow-roll inflation to almost
flatness. In the near future, observations of the microwave background
with the new generation of satellites, MAP and Planck, will determine
with better than 1\% accuracy whether we live in an open universe or
not. It is therefore crucial to know whether inflation can be made
compatible with such a universe. Single-bubble open inflation models
provide a natural scenario for understanding the large scale
homogeneity and isotropy. Furthermore, these inflationary models
generically predict a nearly scale invariant spectrum of density and
gravitational wave perturbations, which could be responsible for the
observed CMB temperature anisotropies. Future observations could then
determine whether these models are compatible with the observed
features of the CMB power spectrum. For that purpose it is necessary
to know the predicted power spectrum with great accuracy. Open models
have a more complicated primordial spectrum of perturbations, with
extra discrete modes and possibly large tensor anisotropies. In order
to constrain those models we have to compute the full spectrum for a
large range of parameters.

In this review we have shown that the simplest single-field models of
open inflation are not only fine tuned, but actually ruled out because
they induce too large tensor anisotropies in the CMB, which is
incompatible with present observations. On the other hand, two-field
models generically do not lead to infinite open universes, as
previously thought, but to an ensemble of very large but finite
inflating `islands'. Each one of these islands will be a quasi-open
universe. We may happen to live in one of those patches, where the
universe {\em appears} to be open. This new effect, semiclassical in
origin, was recently discussed in Ref.~\cite{quasi} where it was found
that many of the present models are in fact quasi-open. This does not
mean that they are not good cosmological models. If the co-moving size
of the inflating islands is sufficiently large, then the resulting
semiclassical anisotropy may be unobservable. We have shown however
that such a component imposes very stringent constraints on the
models. Most of them have a narrow range of parameters for which they
are compatible with observations.

It is perhaps worth mentioning here some alternative proposals (not
single-bubble) for the generation of an open universe in the context
of inflation. First of all, the group of Roma~\cite{Roma} proposed a
model based on higher order gravity that induces bubble nucleation and
later percolation, resulting in a distribution peaked
at $\Omega_0\simeq0.2$. Perhaps the most striking recent results are
those of Hawking-Turok~\cite{HawTur} and Linde~\cite{creation}, who
claim that an open inflationary universe could have been created
directly from the vacuum, without the intermediate de Sitter phase.

\section*{Acknowledgements}

It is a pleasure to thank Andrei Linde, Andrew Liddle, Jaume Garriga
and Xavier Montes, with whom part of the work presented here was done,
for enjoyable discussions on the issues of open inflation. I
acknowledge the use of CMBFAST for the computation of the scalar and
tensor temperature power spectra in Fig.~\ref{fig6}.

\end{document}